\documentclass[10pt]{article}
\usepackage{graphicx}
\usepackage{amsmath}
\usepackage{amssymb}
\usepackage{caption2}
\begin{document}
\begin{center}
{\large {\bf \sc{  Light tetraquark state candidates   }}} \\[2mm]
Zhi-Gang  Wang \footnote{E-mail: zgwang@aliyun.com.  }     \\
 Department of Physics, North China Electric Power University, Baoding 071003, P. R. China
\end{center}

\begin{abstract}
In this article, we study the axialvector-diquark-axialvector-antidiquark type  scalar, axialvector, tensor and vector  $ss\bar{s}\bar{s}$  tetraquark states
with the QCD sum rules.   The predicted mass  $m_{X}=2.08\pm0.12\,\rm{GeV}$ for the axialvector tetraquark state is in excellent agreement with the experimental value $(2062.8 \pm 13.1 \pm 4.2) \,\rm{MeV}$ from the BESIII  collaboration and supports assigning the new $X$ state to be a $ss\bar{s}\bar{s}$ tetraquark state with $J^{PC}=1^{+-}$. The predicted mass $m_{X}=3.08\pm0.11\,\rm{GeV}$  disfavors assigning the $\phi(2170)$ or $Y(2175)$  to be the vector partner of the new $X$ state.  As a byproduct,  we  obtain the masses of the corresponding $qq\bar{q}\bar{q}$ tetraquark states. The light tetraquark states lie in the region about $2\,\rm{GeV}$ rather than $1\,\rm{GeV}$.
 \end{abstract}



\section{Introduction}

Recently, the BESIII  collaboration  studied the process $J/\psi \to \phi \eta \eta^\prime$ and observed a structure $X$ in  the $\phi\eta^\prime$ mass spectrum \cite{BES-2000}. The fitted mass and width are $m_X=(2002.1\pm 27.5 \pm 15.0)\,\rm{MeV}$ and $\Gamma_X=(129 \pm 17 \pm 7)\,\rm{MeV}$ respectively with assumption of the  spin-parity $J^P=1^-$, the corresponding  significance is $5.3\sigma$; while the fitted mass and width are $m_X=((2062.8 \pm 13.1 \pm 4.2)
\,\rm{MeV}$ and $\Gamma_X=(177 \pm 36 \pm 20)\,\rm{MeV}$ respectively with assumption of the  spin-parity  $J^P=1^+$, the corresponding significance is $4.9\sigma$.
The $X$ state was observed in the $\phi\eta^\prime$ decay model rather than in the $\phi\eta$ decay model, they maybe contain a large $ss\bar{s}\bar{s}$ component, in other words, it maybe have a large tetraquark component. In Ref.\cite{Wang-Luo-Liu}, Wang, Luo and Liu assign the $X$ state to be the second radial excitation of the $h_1(1380)$. In Ref.\cite{Cui-etal}, Cui et al assign the $X$ to be  the partner of the tetraquark state $Y(2175)$ with the $J^{PC}=1^{+-}$.

We usually assign  the lowest scalar  nonet  mesons $\{f_0(500),a_0(980),\kappa_0(800),f_0(980) \}$ to be  tetraquark
states,  and assign the higher  scalar nonet mesons
$\{f_0(1370),a_0(1450),K^*_0(1430),f_0(1500) \}$ to be the
conventional ${}^3P_0$ quark-antiquark  states \cite{Close2002,ReviewAmsler2,Maiani-Scalar}.
In Ref.\cite{WangScalarNonet},  we take the nonet scalar mesons  below $1\,\rm{ GeV}$ as the two-quark-tetraquark  mixed states and study their  masses and pole residues  with the  QCD sum rules in details, and observe that the  dominant Fock components of the nonet scalar mesons  below $1\,\rm{ GeV}$ are conventional two-quark states.
The light tetraquark states maybe lie in the region about $2\,\rm{GeV}$ rather than lie in the region about $1\,\rm{GeV}$.

In this article, we take the axialvector diquark operators as the basic constituents to construct the tetraquark current operators to study the scalar ($S$), axialvector ($A$), tensor ($T$) and vector ($V$) tetraquark states with the QCD sum rules, explore the possible assignments of the new $X$ state.  We take  the axialvector diquark operators  as the basic constituents because   the favored configurations from the QCD sum rules are the scalar and axialvector diquark states \cite{WangLDiquark,Dosch-Diquark-1989}, the current operators or quark structures chosen in the present work differ from that in Ref.\cite{Cui-etal} completely.

The article is arranged as follows:  we derive the QCD sum rules for the masses and pole residues of  the $ss\bar{s}\bar{s}$  tetraquark states in section 2; in section 3, we   present the numerical results and discussions; section 4 is reserved for our conclusion.

\section{QCD sum rules for  the  $ss\bar{s}\bar{s}$ tetraquark states}
We write down  the two-point correlation functions $\Pi_{\mu\nu\alpha\beta}(p)$ and $\Pi(p)$ firstly,
\begin{eqnarray}
\Pi_{\mu\nu\alpha\beta}(p)&=&i\int d^4x e^{ip \cdot x} \langle0|T\left\{J_{\mu\nu}(x)J_{\alpha\beta}^{\dagger}(0)\right\}|0\rangle \, , \\
\Pi(p)&=&i\int d^4x e^{ip \cdot x} \langle0|T\left\{J_0(x)J_0^{\dagger}(0)\right\}|0\rangle \, ,
\end{eqnarray}
where $J_{\mu\nu}(x)=J_{2,\mu\nu}(x)$, $J_{1,\mu\nu}(x)$,
\begin{eqnarray}
 J_{2,\mu\nu}(x)&=&\frac{\varepsilon^{ijk}\varepsilon^{imn}}{\sqrt{2}}\Big\{s^{Tj}(x)C\gamma_\mu s^k(x) \bar{s}^{m}(x)\gamma_\nu C \bar{s}^{Tn}(x)+s^{Tj}(x)C\gamma_\nu s^k(x)\bar{s}^m(x)\gamma_\mu C \bar{s}^{Tn}(x) \Big\} \, , \nonumber\\
 J_{1,\mu\nu}(x)&=&\frac{\varepsilon^{ijk}\varepsilon^{imn}}{\sqrt{2}}\Big\{s^{Tj}(x)C\gamma_\mu s^k(x) \bar{s}^{m}(x)\gamma_\nu C \bar{s}^{Tn}(x)-s^{Tj}(x)C\gamma_\nu s^k(x)\bar{s}^m(x)\gamma_\mu C \bar{s}^{Tn}(x) \Big\} \, , \nonumber\\
 J_0(x)&=&\varepsilon^{ijk}\varepsilon^{imn}s^{Tj}(x)C\gamma_\mu s^k(x) \bar{s}^m(x)\gamma^\mu C \bar{s}^{Tn}(x) \, ,
\end{eqnarray}
where the $i$, $j$, $k$, $m$, $n$ are color indexes, the $C$ is the charge conjugation matrix.
 Under charge conjugation transform $\widehat{C}$, the currents $J_{\mu\nu}(x)$ and $J_0(x)$ have the   properties,
\begin{eqnarray}
\widehat{C}\,J_{2,\mu\nu}(x)\,\widehat{C}^{-1}&=&+ \,J_{2,\mu\nu}(x)\, , \nonumber \\
\widehat{C}\,J_{1,\mu\nu}(x)\,\widehat{C}^{-1}&=&- \,J_{1,\mu\nu}(x)\, , \nonumber \\
\widehat{C}\,J_0(x)\,\widehat{C}^{-1}&=& +J_0(x)  \, .
\end{eqnarray}

The doubly-strange diquark operators
\begin{eqnarray}
s^{Tj} C\Gamma s^k&=&\frac{1}{2}\Big(s^{Tj} C\Gamma s^k-s^{Tk} C\Gamma s^j \Big)=\frac{1}{2}\varepsilon^{ijk}s^{Tj} C\Gamma s^k
\end{eqnarray}
with $\Gamma=\gamma_\mu$, $\sigma_{\mu\nu}$ in  color antitriplet $\bar{3}_c$ and
\begin{eqnarray}
s^{Tj} C\Gamma s^k&=&\frac{1}{2}\Big(s^{Tj} C\Gamma s^k+s^{Tk} C\Gamma s^j \Big)
\end{eqnarray}
with $\Gamma=1$, $\gamma_{5}$, $\gamma_{\mu}\gamma_5$ in color sextet  $6_c$ satisfy  Fermi-Dirac statistics. On the other hand, the scattering amplitude for one-gluon exchange  is proportional to
\begin{eqnarray}
\left(\frac{\lambda^a}{2}\right)_{ij}\left(\frac{\lambda^a}{2}\right)_{kl}&=&-\frac{1}{3}\left(\delta_{ij}\delta_{kl}-\delta_{il}\delta_{kj}\right)
+\frac{1}{6}\left(\delta_{ij}\delta_{kl}+\delta_{il}\delta_{kj}\right) \, ,
\end{eqnarray}
where
\begin{eqnarray}
\varepsilon_{mik}\varepsilon_{mjl} &=&\delta_{ij}\delta_{kl}-\delta_{il}\delta_{kj}\, ,
\end{eqnarray}
the $\lambda^a$ is the  Gell-Mann matrix.  The negative sign in front of the antisymmetric  antitriplet $\bar{3}_c$ indicates the interaction
is attractive, which favors  formation of
the diquarks in  color antitriplet.  The positive sign in front of the symmetric sextet $6_c$ indicates
 the interaction  is repulsive,  which  disfavors  formation of
the diquarks in  color sextet. The diquark states which couple potentially to the $s^{Tj} C s^k$, $s^{Tj} C\gamma_5 s^k$ and $s^{Tj} C\gamma_\mu\gamma_5 s^k$ operators  in color sextet  $6_c$ are expected to have larger masses than the diquark states which couple potentially to the $s^{Tj} C\gamma_\mu s^k$  and $s^{Tj} C\sigma_{\mu\nu} s^k$ operators in color antitriplet $\bar{3}_c$.
We prefer the  diquark operators in  color antitriplet $\bar{3}_c$ to the diquark operators in  color sextet  $6_c$ in constructing the tetraquark current operators.
Up to now,  the scalar and axialvector diquark states in color antitriplet $\bar{3}_c$ have been studied with the QCD sum rules \cite{WangLDiquark,Dosch-Diquark-1989}.
In our previous studies, we observed that the pseudoscalar and vector diquark states in color antitriplet $\bar{3}_c$ are not favored configurations, and cannot lead to stable QCD sum rules, which are not included in Ref.\cite{WangLDiquark}.
 The tensor diquark states, which have both $J^P=1^+$ and $1^-$ components,  have not been studied with the QCD sum rules yet. We can draw the conclusion tentatively that the most favored quark configuration is the axialvector diquark operator $\varepsilon^{ijk}s^{Tj} C\gamma_\mu s^k$.
In Ref.\cite{Cui-etal}, Cui et al choose  the pseudoscalar diquark operator in color sextet $6_c$ and vector antidiquark operator in color antisextet $\bar{6}_c$, and  axialvector diquark operator in color antitriplet $\bar{3}_c$ and tensor antidiquark operator  in color triplet $3_c$  to construct the axialvector currents to study the axialvector tetraquark states.  In Ref.\cite{Wang2007NPA}, we choose the color octet-octet type vector four-quark current  to study the $Y(2175)$,
 Fierz rearrangement of this current cannot lead to a  diquark-antidiquark type tensor component.
 In the present work, we choose the axialvector diquark (antidiquark) operators in color antitriplet $\bar{3}_c$ (triplet $3_c$) to construct the tensor current, which is expected to couple potentially to the lowest tetraquark states, to study both the
axialvector and vector tetraquark states.
 The quark configuration in the present work differs completely from that in Ref.\cite{Cui-etal} and Ref.\cite{Wang2007NPA}, it is interesting to study the new quark configuration. Furthermore, the conclusion of the present work differs  completely from that of Ref.\cite{Cui-etal}.

At the hadronic side, we can insert  a complete set of intermediate hadronic states with
the same quantum numbers as the current operators $J_{\mu\nu}(x)$ and $J_0(x)$ into the
correlation functions $\Pi_{\mu\nu\alpha\beta}(p)$ and $\Pi(p)$ to obtain the hadronic representation
\cite{SVZ79,Reinders85}. After isolating the ground state
contributions of the scalar, axialvector, vector and tensor tetraquark states, we get the results,
\begin{eqnarray}
\Pi_{2,\mu\nu\alpha\beta}(p)&=&\frac{\lambda_{ X_T}^2}{m_{X_T}^2-p^2}\left( \frac{\widetilde{g}_{\mu\alpha}\widetilde{g}_{\nu\beta}+\widetilde{g}_{\mu\beta}\widetilde{g}_{\nu\alpha}}{2}-\frac{\widetilde{g}_{\mu\nu}\widetilde{g}_{\alpha\beta}}{3}\right) +\cdots   \nonumber\\
&=&\Pi_{2^+}(p)\left( \frac{\widetilde{g}_{\mu\alpha}\widetilde{g}_{\nu\beta}+\widetilde{g}_{\mu\beta}\widetilde{g}_{\nu\alpha}}{2}-\frac{\widetilde{g}_{\mu\nu}\widetilde{g}_{\alpha\beta}}{3}\right) +\cdots \, ,
\end{eqnarray}

\begin{eqnarray}
\Pi_{1,\mu\nu\alpha\beta}(p)&=&\frac{\widetilde{\lambda}_{ X_A}^2}{m_{X_A}^2-p^2}\left(p^2g_{\mu\alpha}g_{\nu\beta} -p^2g_{\mu\beta}g_{\nu\alpha} -g_{\mu\alpha}p_{\nu}p_{\beta}-g_{\nu\beta}p_{\mu}p_{\alpha}+g_{\mu\beta}p_{\nu}p_{\alpha}+g_{\nu\alpha}p_{\mu}p_{\beta}\right) \nonumber\\
&&+\frac{\widetilde{\lambda}_{ X_V}^2}{m_{X_V}^2-p^2}\left( -g_{\mu\alpha}p_{\nu}p_{\beta}-g_{\nu\beta}p_{\mu}p_{\alpha}+g_{\mu\beta}p_{\nu}p_{\alpha}+g_{\nu\alpha}p_{\mu}p_{\beta}\right) +\cdots \nonumber\\
&=&\Pi_{1^+}(p^2)\left(p^2g_{\mu\alpha}g_{\nu\beta} -p^2g_{\mu\beta}g_{\nu\alpha} -g_{\mu\alpha}p_{\nu}p_{\beta}-g_{\nu\beta}p_{\mu}p_{\alpha}+g_{\mu\beta}p_{\nu}p_{\alpha}+g_{\nu\alpha}p_{\mu}p_{\beta}\right) \nonumber\\
&&+\Pi_{1^-}(p^2)\left( -g_{\mu\alpha}p_{\nu}p_{\beta}-g_{\nu\beta}p_{\mu}p_{\alpha}+g_{\mu\beta}p_{\nu}p_{\alpha}+g_{\nu\alpha}p_{\mu}p_{\beta}\right) \, ,
\end{eqnarray}

\begin{eqnarray}
\Pi(p)&=&\Pi_{0^+}(p^2)=\frac{\lambda_{ X_S}^2}{m_{X_S}^2-p^2} +\cdots \, \, ,
\end{eqnarray}
where $\widetilde{g}_{\mu\nu}=g_{\mu\nu}-\frac{p_\mu p_\nu}{p^2}$, the subscripts $2^+$, $1^+$, $1^-$ and $0^+$ denote the spin-parity $J^P$ of the corresponding tetraquark states.
The pole residues  $\lambda_{X}$ and $\widetilde{\lambda}_{X}$  are defined by
\begin{eqnarray}
 \langle 0|J_{2,\mu\nu}(0)|X_{T}(p)\rangle &=& \lambda_{X_T} \, \varepsilon_{\mu\nu} \, , \nonumber\\
  \langle 0|J_{1,\mu\nu}(0)|X_{A}(p)\rangle &=& \widetilde{\lambda}_{X_A} \, \varepsilon_{\mu\nu\alpha\beta} \, \varepsilon^{\alpha}p^{\beta}\, , \nonumber\\
 \langle 0|J_{1,\mu\nu}(0)|X_{V}(p)\rangle &=& \widetilde{\lambda}_{X_V} \left(\varepsilon_{\mu}p_{\nu}-\varepsilon_{\nu}p_{\mu} \right)\, , \nonumber\\
   \langle 0|J_0(0)|X_{S}(p)\rangle &=& \lambda_{X_S} \, ,
\end{eqnarray}
where the $\varepsilon_{\mu\nu}$ and $\varepsilon_\mu$ are the polarization vectors of the  tetraquark states.

Now we contract the $s$ quarks in the correlation functions with Wick theorem,  there are four $s$-quark propagators,  if two $s$-quark lines emit a gluon by itself and the other two $s$-quark lines contribute  a quark pair by itself, we obtain a operator $GG\bar{s}s\bar{s}s$, which is  of order ${\mathcal{O}}(\alpha_s^k)$ with $k=1$ and of dimension $10$. In this article,  we take into account the vacuum condensates up  to dimension $10$ and $k\leq 1$ in a consistent way. For the technical details, one can consult Refs.\cite{WangScalarNonet,WangHuangtao-2014}.
Once  the analytical expressions of the QCD spectral densities are obtained,  we take the
quark-hadron duality below the continuum thresholds  $s_0$ and perform Borel transform  with respect to
the variable $P^2=-p^2$ to obtain   the  QCD sum rules:
\begin{eqnarray}\label{QCDSR}
\lambda^2_{X}\, \exp\left(-\frac{m^2_{X}}{T^2}\right)= \int_{0}^{s_0} ds\, \rho(s) \, \exp\left(-\frac{s}{T^2}\right) \, ,
\end{eqnarray}
where $\rho(s)=\rho_S(s)$, $\rho_A(s)$, $\rho_V(s)$ and $\rho_T(s)$,
\begin{eqnarray}
\rho_{S}(s)&=& \frac{s^4}{3840\pi^6}   -\frac{13s\,m_s\langle\bar{s}g_{s}\sigma Gs\rangle}{384\pi^4}
+\frac{2s\langle\bar{s}s\rangle^2}{3\pi^2} -\frac{17\langle\bar{s}s\rangle \langle\bar{s}g_{s}\sigma Gs\rangle}{48\pi^2}+\frac{s^2}{192\pi^4}\langle\frac{\alpha_{s}GG}{\pi}\rangle  \nonumber\\
&& +\frac{19m_s\langle\bar{s}s\rangle}{96\pi^2} \langle\frac{\alpha_{s}GG}{\pi}\rangle-\frac{16m_s\langle\bar{s}s\rangle^3}{3}\delta(s)+\frac{\langle\bar{s}g_{s}\sigma Gs\rangle^2}{192\pi^2}\delta(s) -\frac{\langle\bar{s}s\rangle^2}{24}\langle\frac{\alpha_{s}GG}{\pi}\rangle\delta(s)  \, ,
\end{eqnarray}

\begin{eqnarray}
\rho_{A}(s)&=& \frac{s^4}{11520\pi^6}-\frac{s^2\,m_s\langle\bar{s}s\rangle}{12\pi^4}
+\frac{s\,m_s\langle\bar{s}g_{s}\sigma Gs\rangle}{9\pi^4} +\frac{4s\,\langle\bar{s}s\rangle^2}{9\pi^2}
-\frac{5\langle\bar{s}s\rangle \langle\bar{s}g_{s}\sigma Gs\rangle}{18\pi^2}  \nonumber\\
&&-\frac{s^2}{2304\pi^4}\langle\frac{\alpha_{s}GG}{\pi}\rangle+\frac{3m_s\langle\bar{s}s\rangle}{64\pi^2}\langle\frac{\alpha_{s}GG}{\pi}\rangle-\frac{32m_s\langle\bar{s}s\rangle^3}{9}\delta(s) -\frac{2\langle\bar{s}s\rangle^2}{27}\langle\frac{\alpha_{s}GG}{\pi}\rangle\delta(s) \, ,
\end{eqnarray}

\begin{eqnarray}
\rho_{V}(s)&=& \frac{s^4}{11520\pi^6}+\frac{s^2\,m_s\langle\bar{s}s\rangle}{12\pi^4}-\frac{7s\,m_s\langle\bar{s}g_{s}\sigma Gs\rangle}{72\pi^4}
-\frac{2s\langle\bar{s}s\rangle^2}{9\pi^2} +\frac{5\langle\bar{s}s\rangle \langle\bar{s}g_{s}\sigma Gs\rangle}{18\pi^2} \nonumber\\
&&+\frac{s^2}{768\pi^4}\langle\frac{\alpha_{s}GG}{\pi}\rangle-\frac{79m_s\langle\bar{s}s\rangle}{1728\pi^2}\langle\frac{\alpha_{s}GG}{\pi}\rangle +\frac{16m_s\langle\bar{s}s\rangle^3}{9}\delta(s) \nonumber\\
&&-\frac{2\langle\bar{s}s\rangle^2}{81}\langle\frac{\alpha_{s}GG}{\pi}\rangle\delta(s) -\frac{\langle\bar{s}g_{s}\sigma Gs\rangle^2}{18\pi^2}\delta(s) \, ,
\end{eqnarray}

\begin{eqnarray}
\rho_T(s)&=&\frac{s^4}{5376\pi^6}-\frac{3s^2\,m_s\langle\bar{s}s\rangle}{20\pi^4} +\frac{29s\,m_s\langle\bar{s}g_{s}\sigma Gs\rangle}{96\pi^4}
+\frac{8s\langle\bar{s}s\rangle^2}{9\pi^2}-\frac{37\langle\bar{s}s\rangle \langle\bar{s}g_{s}\sigma Gs\rangle}{48\pi^2} \nonumber\\
&&-\frac{11s^2}{1920\pi^4}\langle\frac{\alpha_{s}GG}{\pi}\rangle+\frac{43m_s\langle\bar{s}s\rangle}{864\pi^2}\langle\frac{\alpha_{s}GG}{\pi}\rangle
-\frac{64m_s\langle\bar{s}s\rangle^3}{9}\delta(s)  -\frac{4\langle\bar{s}s\rangle^2}{27}\langle\frac{\alpha_{s}GG}{\pi}\rangle\delta(s) \, , \nonumber\\
\end{eqnarray}
and $\lambda_{X_{A/V}}=m_{X_{A/V}}\widetilde{\lambda}_{X_{A/V}}$.

We derive Eq.\eqref{QCDSR} with respect to  $\tau=\frac{1}{T^2}$, then  obtain the QCD sum rules for  the masses of the tetraquark states through a fraction,
 \begin{eqnarray}\label{mass-QCDSR}
m^2_{X}&=& -\frac{\int_{0}^{s_0} ds\frac{d}{d \tau}\rho(s)\exp\left(-\tau s \right)}{\int_{0}^{s_0} ds \rho(s)\exp\left(-\tau s\right)}\, .
\end{eqnarray}

\section{Numerical results and discussions}
We take  the standard values of the vacuum condensates $\langle
\bar{q}q \rangle=-(0.24\pm 0.01\, \rm{GeV})^3$,   $\langle
\bar{q}g_s\sigma G q \rangle=m_0^2\langle \bar{q}q \rangle$,
$m_0^2=(0.8 \pm 0.1)\,\rm{GeV}^2$, $\langle\bar{s}s \rangle=(0.8\pm0.1)\langle\bar{q}q \rangle$, $\langle\bar{s}g_s\sigma G s \rangle=m_0^2\langle \bar{s}s \rangle$,  $\langle \frac{\alpha_s
GG}{\pi}\rangle=(0.012\pm0.004)\,\rm{GeV}^4 $    at the energy scale  $\mu=1\, \rm{GeV}$
\cite{SVZ79,Reinders85,Colangelo-Review}, and choose the $\overline{MS}$ mass $m_s(\mu=2\,\rm{GeV})=0.095\pm 0.005\,\rm{GeV}$ from the Particle Data Group \cite{PDG}, and evolve the $s$-quark mass to the energy scale $\mu=1\,\rm{GeV}$ with the renormalization group equation,  furthermore, we neglect the small $u$ and $d$ quark masses.

\begin{figure}
 \centering
  \includegraphics[totalheight=5cm,width=7cm]{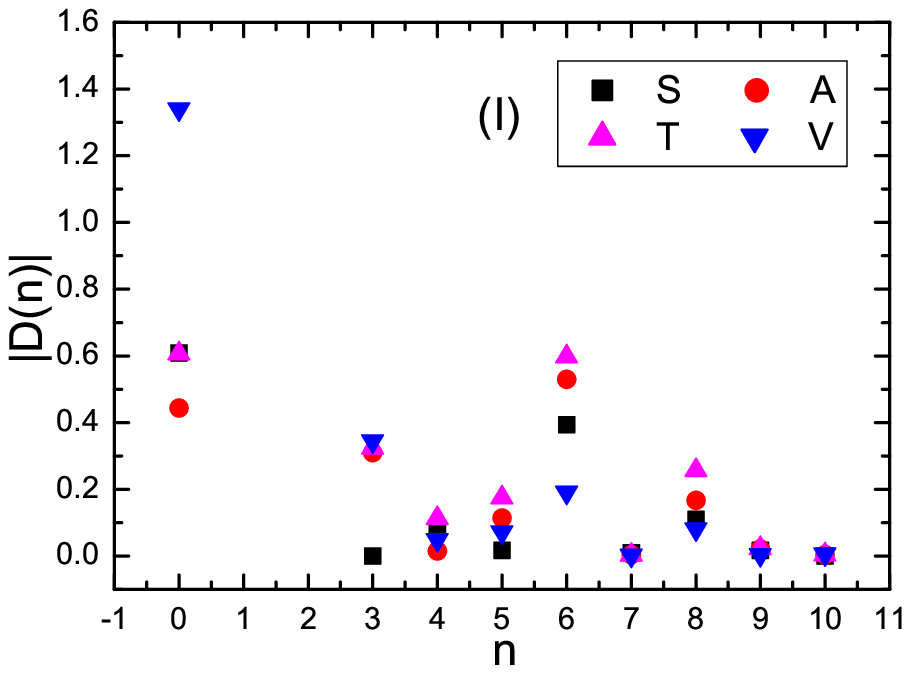}
  \includegraphics[totalheight=5cm,width=7cm]{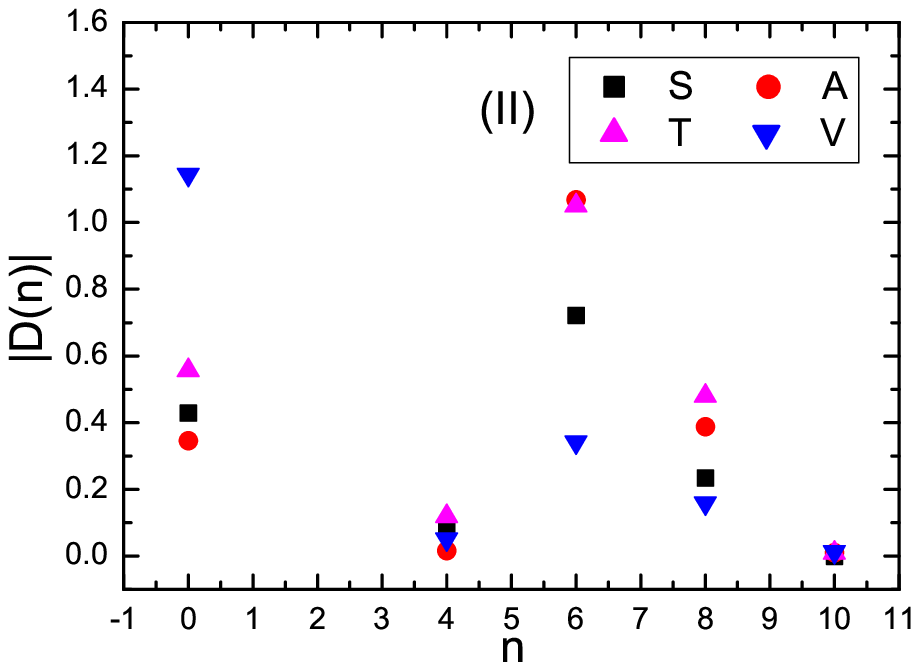}
     \caption{The absolute contributions of the vacuum condensates of dimension $n$ for the central values of the  input  parameters  in the operator product expansion, where the $S$, $A$, $T$ and $V$ denote the  scalar, axialvector, tensor and vector tetraquark states, respectively, the (I) and (II) denote the $ss\bar{s}\bar{s}$ and $qq\bar{q}\bar{q}$ quark constituents, respectively.}\label{OPE}
\end{figure}

 We choose suitable Borel parameters and continuum threshold parameters to warrant the pole contributions (PC) are larger than $40\%$, i.e.
\begin{eqnarray}
\text{PC}&=&\frac{\int_{0}^{s_{0}}ds\,\rho\left(s\right)\exp\left(-\frac{s}{T^{2}}\right)} {\int_{0}^{\infty}ds\,\rho\left(s\right)\exp\left(-\frac{s}{T^{2}}\right)}\geq 40\%\ ,
\end{eqnarray}
and convergence of the operator product expansion. The contributions of the vacuum condensates $D(n)$
in the operator product expansion are defined by,
\begin{eqnarray}
D(n)&=&\frac{\int_{0}^{s_{0}}ds\,\rho_{n}(s)\exp\left(-\frac{s}{T^{2}}\right)}
{\int_{0}^{s_{0}}ds\,\rho\left(s\right)\exp\left(-\frac{s}{T^{2}}\right)}\ ,
\end{eqnarray}
where the subscript $n$ in the QCD spectral density $\rho_{n}(s)$ denotes the dimension of the vacuum condensates. We choose the values $|D(10)|\sim 1\%$ to warrant the convergence of the operator product expansion.
In Table \ref{mass-residue}, we present the ideal Borel parameters, continuum threshold parameters, pole contributions and contributions of the vacuum condensates of dimension $10$.
In Fig.\ref{OPE}, we plot the absolute  contributions of the vacuum condensates of dimension $n$ for the central values of the  input  parameters   in the operator product expansion. Although in some cases,  the contributions of the perturbative terms $D(0)$ are not  the dominant contributions, the contributions of the vacuum condensates of dimensions $6$ and $8$ are very large,  the hierarchy $|D(6)|\gg |D(8)|$ warrants the good convergent behavior of the operator product expansion, furthermore, the contributions $D(7)$, $D(9)$ and $D(10)$ are very small.
From Table \ref{mass-residue} and Fig.\ref{OPE}, we can see that the pole dominance is well satisfied and the operator product expansion is well convergent, we expect to make reliable predictions.

\begin{table}
\begin{center}
\begin{tabular}{|c|c|c|c|c|c|c|c|}\hline\hline
                       &$T^2 (\rm{GeV}^2)$ &$\sqrt{s_0}(\rm{GeV})$  &pole         &$|D(10)|$ &$m_{X}(\rm{GeV})$  &$\lambda_{X}(10^{-2}\rm{GeV}^5)$\\ \hline

$ss\bar{s}\bar{s}_S$   &$1.4-1.8$          &$2.65\pm0.10$           &$(40-73)\%$  &$\ll1\%$   &$2.08\pm0.13$      &$2.73\pm 0.56$                    \\ \hline

$ss\bar{s}\bar{s}_A$   &$1.5-1.9$          &$2.65\pm0.10$           &$(41-72)\%$  &$<1\%$     &$2.08\pm0.12$      &$1.87\pm 0.34$                    \\ \hline

$ss\bar{s}\bar{s}_T$   &$1.5-1.9$          &$2.75\pm0.10$           &$(41-72)\%$  &$<1\%$     &$2.22\pm0.11$      &$3.02\pm 0.53$                    \\ \hline

$ss\bar{s}\bar{s}_V$   &$2.1-2.7$          &$3.60\pm0.10$           &$(42-73)\%$  &$\leq1\%$  &$3.08\pm0.11$      &$6.47\pm 1.07$                    \\ \hline

$qq\bar{q}\bar{q}_S$   &$1.2-1.6$          &$2.40\pm0.10$           &$(40-76)\%$  &$\ll1\%$   &$1.86\pm0.11$      &$1.95\pm 0.38$                    \\ \hline

$qq\bar{q}\bar{q}_A$   &$1.3-1.7$          &$2.40\pm0.10$           &$(40-73)\%$  &$\leq1\%$  &$1.87\pm0.10$      &$1.30\pm 0.22$                    \\ \hline

$qq\bar{q}\bar{q}_T$   &$1.4-1.8$          &$2.65\pm0.10$           &$(42-74)\%$  &$\leq1\%$  &$2.13\pm0.10$      &$2.58\pm 0.42$                    \\ \hline

$qq\bar{q}\bar{q}_V$   &$1.9-2.5$          &$3.40\pm0.10$           &$(41-74)\%$  &$\leq2\%$  &$2.86\pm0.11$      &$4.94\pm 0.93$                    \\ \hline

 \hline
\end{tabular}
\end{center}
\caption{ The Borel parameters, continuum threshold parameters, pole contributions,   contributions of the vacuum condensates of dimension $10$, masses and pole residues of the tetraquark states, where the subscripts $S$, $A$, $T$ and $V$ denote the scalar, axialvector, tensor and vector tetraquark states, respectively. }\label{mass-residue}
\end{table}

\begin{figure}
 \centering
  \includegraphics[totalheight=5cm,width=7cm]{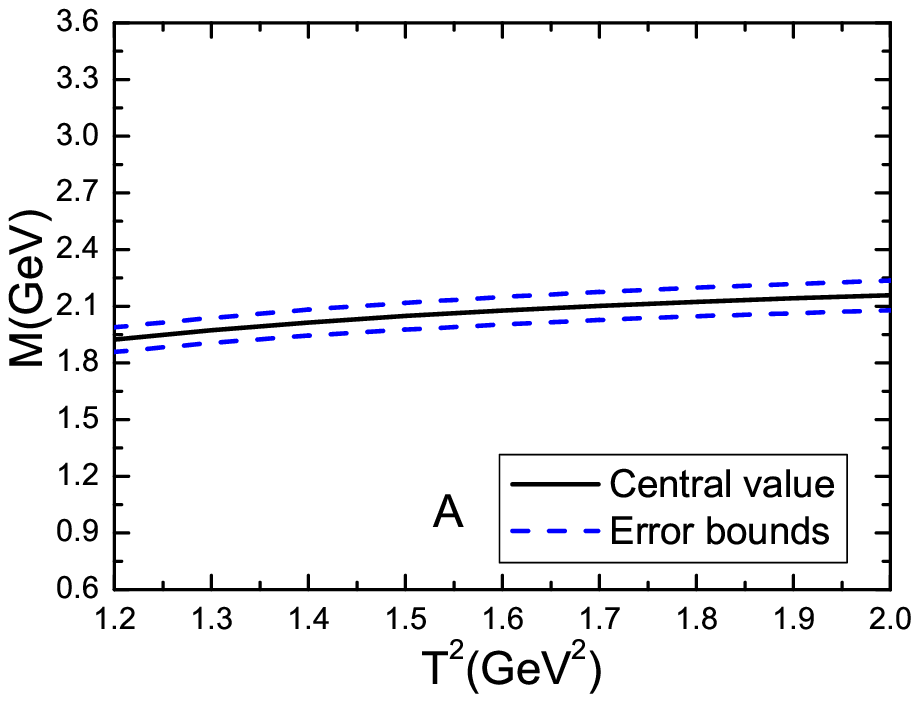}
  \includegraphics[totalheight=5cm,width=7cm]{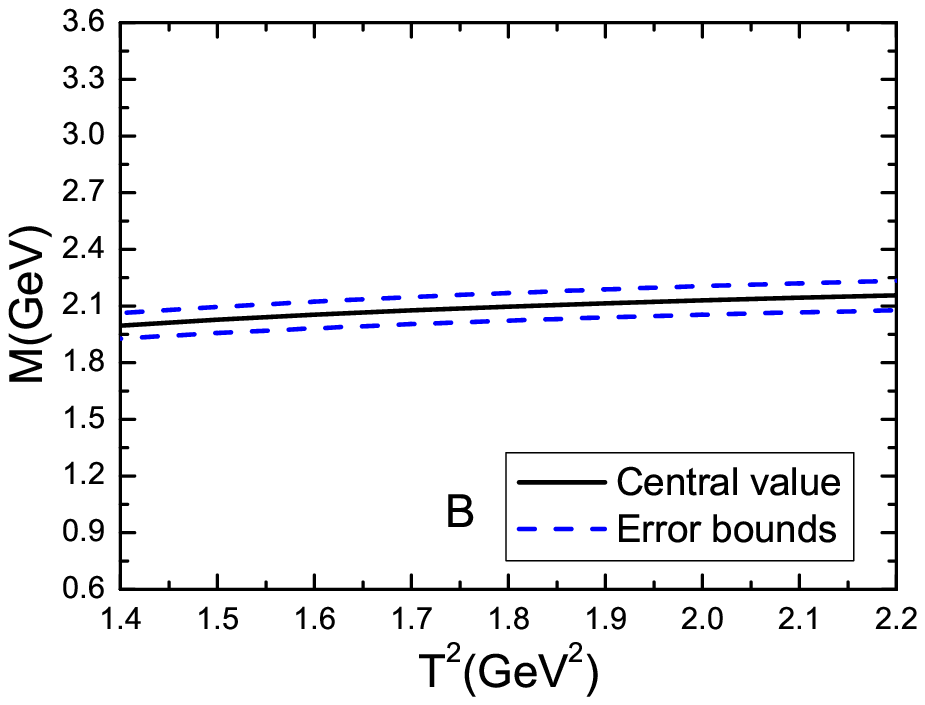}
  \includegraphics[totalheight=5cm,width=7cm]{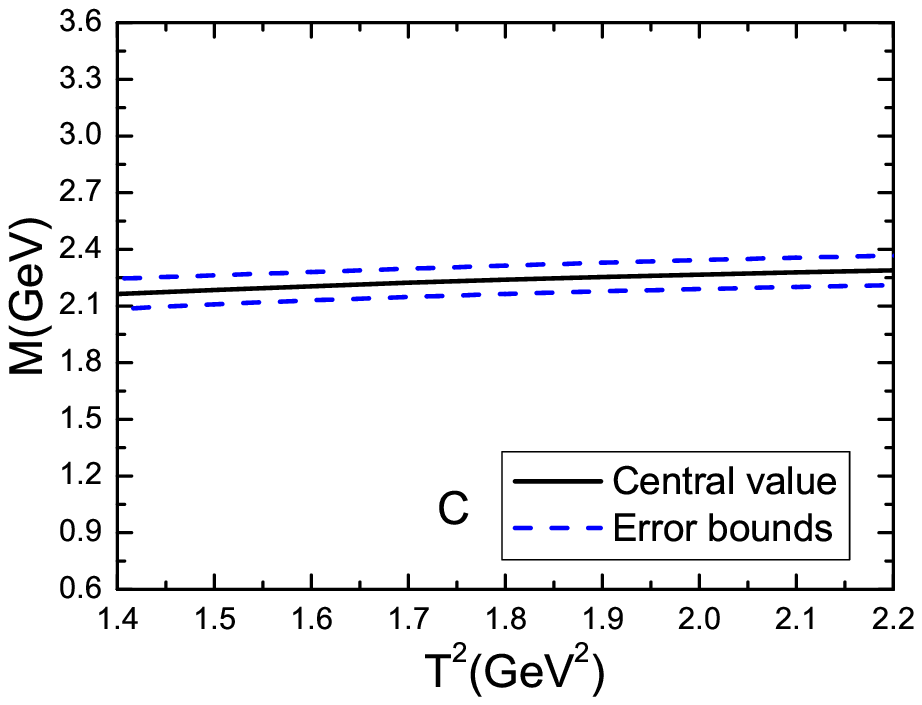}
 \includegraphics[totalheight=5cm,width=7cm]{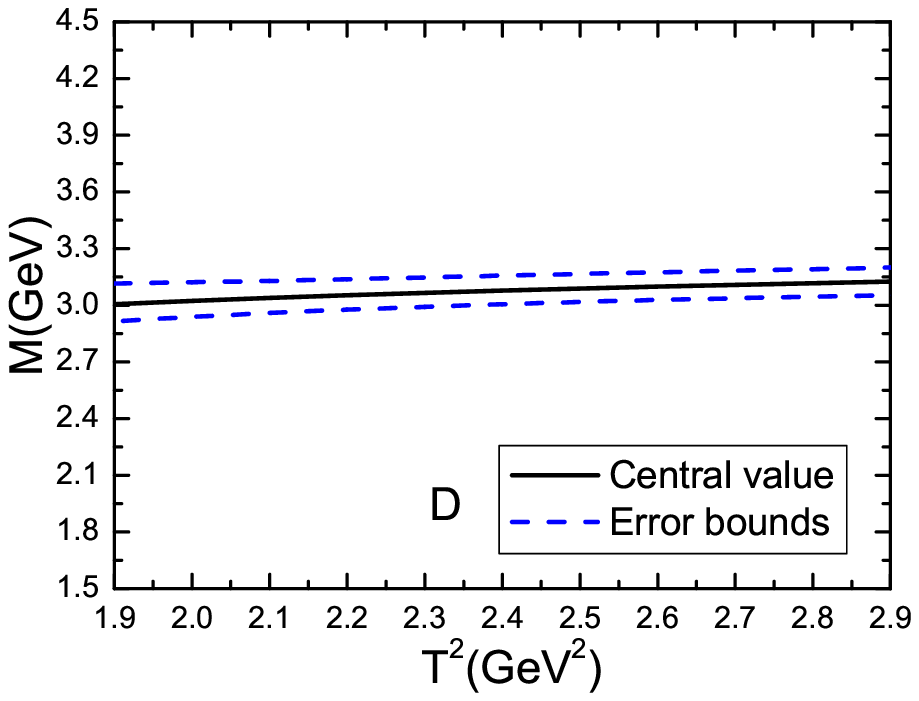}
     \caption{The masses  with variations of the  Borel  parameters  $T^2$, where the $A$, $B$, $C$ and $D$ denote the  scalar, axialvector, tensor and vector tetraquark states, respectively.}\label{mass-ssss}
\end{figure}

We take into account all uncertainties of the input parameters, and obtain the values of the masses and pole residues of
 the   $ss\bar{s}\bar{s}$ tetraquark states, which are  shown explicitly in Fig.\ref{mass-ssss} and Table \ref{mass-residue}. In this article, we have assumed that the energy gaps between the ground state and the first radial state is about $0.6\,\rm{GeV}$ \cite{Wang-Energy-gap}.  In Fig.\ref{mass-ssss}, we plot the masses of the scalar, axialvector, tensor and vector $ss\bar{s}\bar{s}$ tetraquark states with variations of the Borel parameters at larger regions than the Borel windows shown in Table \ref{mass-residue}.  From the figure, we can see that there appear platforms in the Borel windows.

From Table \ref{mass-residue}, we can see that the uncertainties of the masses $\delta M_X$ are small, while the uncertainties of the pole residues $\delta \lambda_X$ are large, for example,
$\frac{\delta M_X}{M_X}=6\%$ and $\frac{\delta \lambda_X}{\lambda_X}=21\%$ for the scalar $ss\bar{s}\bar{s}$ tetraquark state. We obtain the tetraquark masses from a fraction, see Eq.\eqref{mass-QCDSR}, the uncertainties  originate from the input parameters  in the numerator and  denominator are almost canceled out with each other, so the net uncertainties of the tetraquark masses are very small. In this article, we have neglected the perturbative $\mathcal{O}(\alpha_s)$ corrections.
For the traditional two-quark light  mesons, the perturbative $\mathcal{O}(\alpha_s)$ corrections amount to multiplying the perturbative terms with a factor
$1+\frac{11}{3}\frac{\alpha_s}{\pi}$ for the $J^{PC}=0^{+-}$, $0^{++}$ mesons,
$1+\frac{\alpha_s}{\pi}$ for the  $J^{PC}=1^{--}$, $1^{++}$, $1^{+-}$ mesons, and
$1-\frac{\alpha_s}{\pi}$ for the $J^{PC}=2^{++}$ mesons \cite{Reinders85}.   Now we estimate the possible uncertainties due to neglecting the  perturbative $\mathcal{O}(\alpha_s)$ corrections by multiplying
the perturbative terms with a factor  $1+(-1\sim 4)\frac{\alpha_s}{\pi}$. The additional  uncertainties $\delta M_X$ and $\delta \lambda_X$ are shown in Table \ref{mass-afs}. From  the Table, we can see again that the  uncertainties of the mass $\delta M_X$ are small, while the uncertainties of the pole residues $\delta \lambda_X$ are large, for example,
$\frac{\delta M_X}{M_X}={}^{+2\%}_{-1\%}$ and $\frac{\delta \lambda_X}{\lambda_X}={}^{+23\%}_{-7\%}$ for the scalar $ss\bar{s}\bar{s}$ tetraquark state.
  In the QCD sum rules for the $X$, $Y$, $Z$ states, which are excellent candidates for the compact tetraquark states or loosely bound molecular states, the uncertainties of the masses are less than or about $6\%$ \cite{Nielsen-Review}. Ref.\cite{Nielsen-Review} is the most recent review. 

The predicted mass $m_{X}=2.08\pm0.12\,\rm{GeV}$ for the axialvector tetraquark state is in excellent agreement with the experimental value $(2062.8 \pm 13.1 \pm 4.2)
\,\rm{MeV}$ from the BESIII  collaboration \cite{BES-2000}, which supports assigning the new $X$ state to be an  axialvector-diquark-axialvector-antidiquark type  $ss\bar{s}\bar{s}$ tetraquark state. The predicted mass $m_{X}=3.08\pm0.11\,\rm{GeV}$ for the vector tetraquark state lies above the experimental value of the mass of the $\phi(2170)$ or $Y(2175)$, $m_{\phi}=2188\pm10\,\rm{MeV}$, from the Particle Data Group, and disfavors assigning the $\phi(2170)$ or $Y(2175)$ to be vector partner of the new $X$ state. If the $\phi(2170)$ have tetraquark component, it maybe have color octet-octet component \cite{Wang2007NPA}.
As a byproduct, we  obtain the masses and pole residues of the corresponding $qq\bar{q}\bar{q}$ tetraquark states, which are shown in Table \ref{mass-residue}.
The present predictions can be confronted to the experimental data in the future.

Now we perform Fierz rearrangement to the  currents  both in the color and Dirac-spinor  spaces,
\begin{eqnarray}
J_0 &=& 2 \bar{s}s\,\bar{s}s+2\bar{s}i\gamma_5s\,\bar{s}i\gamma_5s+ \bar{s}\gamma_{\alpha} s\,\bar{s}\gamma^{\alpha}s- \bar{s}\gamma_{\alpha}\gamma_5 s\,\bar{s}\gamma^{\alpha}\gamma_5s  \, , \nonumber\\
J_{1,\mu\nu} &=&\sqrt{2}\Big\{\,i\bar{s}s\, \bar{s}\sigma_{\mu\nu}s -\bar{s}\sigma_{\mu\nu}\gamma_5s\,\bar{s}i\gamma_5s   +i\varepsilon_{\mu\nu\alpha\beta}\bar{s}\gamma^\alpha\gamma_5s\, \bar{s}\gamma^\beta s\,\Big\} \, , \nonumber\\
J_{2,\mu\nu} &=&\frac{1}{\sqrt{2}}\Big\{\,2 \bar{s}\gamma_\mu\gamma_5s\, \bar{s}\gamma_\nu\gamma_5s -2\bar{s}\gamma_\mu s\, \bar{s}\gamma_\nu s
   +2g^{\alpha\beta} \bar{s}\sigma_{\mu\alpha}s\, \bar{s}\sigma_{\nu\beta}s+g_{\mu\nu}\Big( \bar{s}s\,\bar{s}s  \nonumber\\
 &&+\bar{s}i\gamma_5s\,\bar{s}i\gamma_5s+\bar{s}\gamma_{\alpha} s\,\bar{s}\gamma^{\alpha}s-\bar{s}\gamma_{\alpha}\gamma_5 s\,\bar{s}\gamma^{\alpha}\gamma_5s-\frac{1}{2}\bar{s}\sigma_{\alpha\beta} s\,\bar{s}\sigma^{\alpha\beta}s \Big) \Big\} \, .
\end{eqnarray}
The diquark-antidiquark type currents  can be re-arranged into  currents as special superpositions of color singlet-singlet type currents, which  couple potentially
 to the meson-meson pairs or molecular states,  the diquark-antidiquark type tetraquark
states can be taken as special superpositions of  meson-meson pairs, and embodies the net effects. The decays to their components are Okubo-Zweig-Iizuka supper-allowed, we can search for those tetraquark states in the decays,
\begin{eqnarray}
X_{S} &\to& \eta^\prime \eta^\prime\, ,\,\, f_0(980) f_0(980)\, ,\,\,\phi(1020)\phi(1020)\, , \nonumber\\
X_{A/V} &\to& f_0(980) h_1(1380)\, ,\,\, \phi(1020) \eta^\prime\, ,\,\,\phi(1020)\phi(1020)\, , \nonumber\\
X_{T} &\to&\eta^\prime \eta^\prime\, ,\,\, f_0(980) f_0(980)\, ,\,\,\phi(1020)\phi(1020)\, .
\end{eqnarray}

\begin{table}
\begin{center}
\begin{tabular}{|c|c|c|c|c|c|c|c|}\hline\hline
                                 &$\delta m_{X}(\rm{GeV})$  &$\delta \lambda_{X}(10^{-2}\rm{GeV}^5)$\\ \hline

$ss\bar{s}\bar{s}_S$             &${}^{+0.04}_{-0.02}$      &${}^{+0.64}_{-0.18}$          \\ \hline

$ss\bar{s}\bar{s}_A$             &${}^{+0.03}_{-0.02}$      &${}^{+0.33}_{-0.09}$           \\ \hline

$ss\bar{s}\bar{s}_T$             &${}^{+0.03}_{-0.01}$      &${}^{+0.63}_{-0.18}$           \\ \hline

$ss\bar{s}\bar{s}_V$             &${}^{+0.03}_{-0.06}$      &${}^{+1.62}_{-0.45}$          \\ \hline

$qq\bar{q}\bar{q}_S$             &${}^{+0.04}_{-0.01}$      &${}^{+0.35}_{-0.10}$        \\ \hline

$qq\bar{q}\bar{q}_A$             &${}^{+0.03}_{-0.01}$      &${}^{+0.18}_{-0.05}$                   \\ \hline

$qq\bar{q}\bar{q}_T$             &${}^{+0.03}_{-0.01}$      &${}^{+0.51}_{-0.14}$                    \\ \hline

$qq\bar{q}\bar{q}_V$             &${}^{+0.02}_{-0.02}$      &${}^{+1.27}_{-0.37}$                    \\ \hline

 \hline
\end{tabular}
\end{center}
\caption{ The possible uncertainties induced by the perturbative $\mathcal{O}(\alpha_s)$ corrections, where the subscripts $S$, $A$, $T$ and $V$ denote the scalar, axialvector, tensor and vector tetraquark states, respectively. }\label{mass-afs}
\end{table}

\section{Conclusion}
In this article, we construct the axialvector-diquark-axialvector-antidiquark type  currents to interpolate   the scalar, axialvector, tensor and vector  $ss\bar{s}\bar{s}$  tetraquark states, then  calculate the contributions of the vacuum condensates up to dimension-10  in the operator product expansion,
and obtain the QCD sum rules for the masses and pole residues of those tetraquark states.   The predicted mass  $m_{X}=2.08\pm0.12\,\rm{GeV}$ for the axialvector tetraquark state is in excellent agreement with the experimental value, $m_X=(2062.8 \pm 13.1 \pm 4.2)
\,\rm{MeV}$, from the BESIII  collaboration and supports assigning the new $X$ state to be an axialvector-diquark-axialvector-antidiquark type  $ss\bar{s}\bar{s}$ tetraquark state. The predicted mass $m_{X}=3.08\pm0.11\,\rm{GeV}$ for the vector tetraquark state lies above the experimental value of the mass of the $\phi(2170)$, $m_{\phi}=2188\pm10\,\rm{MeV}$, from the Particle Data Group,  and disfavors assigning the $\phi(2170)$ to be the vector partner of the new $X$ state.  As a byproduct,  we also obtain the masses and pole residues of the corresponding $qq\bar{q}\bar{q}$ tetraquark states. The present predictions can be confronted to the experimental data in the future.

\section*{Acknowledgements}
This  work is supported by National Natural Science Foundation, Grant Number  11775079.

\end{document}